\documentclass[a4paper, 12pt]{article}
\pdfoutput=1      
\usepackage{authblk} 
\usepackage[top=25truemm,bottom=25truemm,left=20truemm,right=20truemm]{geometry}
\usepackage{hyperref}
\usepackage{setspace}
\usepackage{amsmath,amssymb,mathrsfs,amsbsy,latexsym,amsfonts,amsthm}
\usepackage{fancyhdr}
\usepackage[Symbol]{upgreek}

\usepackage{graphicx}         
\usepackage{color}
\usepackage{slashed}
\usepackage{tensor}
\usepackage{comment}
\usepackage{braket}
\usepackage{ascmac}
\usepackage{here}
\usepackage{bm}

\newcommand{\pc}{\pi_{(c)}}
\newcommand{\bpc}{\bar{\pi}_{(c)}}
\newcommand{\intx}{\int \!\! d^3 x}

\newcommand{\intk}{\int \!\! \frac{d^3 k}{(2\pi)^3}}
\newcommand{\intkw}{\int \!\! \frac{d^3 k}{(2\pi)^3(2\omega)}}
\newcommand{\intp}{\int \!\! \frac{d^3 p}{(2\pi)^3(2 E_p)}}

\newcommand{\vx}{\vec{x}}
\newcommand{\vy}{\vec{y}}
\newcommand{\vk}{\vec{k}}
\newcommand{\vkp}{\vec{k'}}
\newcommand{\vp}{\vec{p}}

\newcommand{\kx}{\vec{k}\cdot \vec{x}}

\newcommand{\nn}{\nonumber\\}
\newcommand{\outin}{\tensor[_{out}]{\braket{\alpha|\beta}}{_{in}}}
\newcommand{\outfree}[1]{\tensor[_0]{\bra{#1}}{}}

\newcommand{\be}{\begin{equation}}
\newcommand{\ee}{\end{equation}}
\newcommand{\ba}[1]{\begin{align}#1\end{align}}

\makeatletter

\@addtoreset{equation}{section}
\makeatother

\begin{document}
	\setlength{\abovedisplayskip}{10pt}
	\setlength{\belowdisplayskip}{10pt}

\title{
	\bf \LARGE 
	Dressed states from gauge invariance
	\vskip 0.5cm
}
\author[1]{
	Hayato~Hirai\thanks{\tt hirai@het.phys.sci.osaka-u.ac.jp}
}
\author[1,2]{
	Sotaro~Sugishita\thanks{\tt sotaro.s@uky.edu}
	\vspace{5mm}
}
\affil[1]{\it\normalsize Department of Physics, Osaka University, Toyonaka, Osaka, 560-0043, Japan}
\affil[2]{\it\normalsize Department of Physics and Astronomy, University of Kentucky, Lexington, KY 40506, USA}
\setcounter{Maxaffil}{0}
\date{}

\maketitle
\thispagestyle{fancy}
\renewcommand{\headrulewidth}{0pt}

\begin{abstract}
	The dressed state formalism enables us to define the infrared finite $S$-matrix for QED. In the formalism, asymptotic charged states are dressed by clouds of photons. The dressed asymptotic states are originally obtained by solving the dynamics of the asymptotic Hamiltonian in the far past or future region. 
	However, 
	there was an argument that the obtained dressed states are not gauge invariant. We resolve the problem by imposing a correct gauge invariant condition. 
	We show that the dressed states can be obtained just by requiring the gauge invariance of asymptotic states. In other words, Gauss's law naturally leads to proper asymptotic states for the infrared finite $S$-matrix. 
	We also discuss the relation between the dressed state formalism and the asymptotic symmetry for QED.  
\end{abstract}

\newpage
\setcounter{tocdepth}{2}
\tableofcontents

\newpage

\setlength{\parskip}{5pt} 
\section{Introduction and summary}\label{sec:intro}

\subsection{Introduction}\label{subsec:intro}

$S$-matrix is a fundamental quantity of quantum field theories in Minkowski spacetime. There is a systematic way to compute it perturbatively by the Feynman rules. 
However, we then often encounter infrared (IR) divergences for theories with massless particles. 
A famous example is quantum electrodynamics (QED). Virtual photons with small energy cause divergences of loop diagrams. 
This problem can be avoided by considering the total cross-section of various processes including the emission of real soft photons \cite{Bloch:1937pw, Yennie:1961ad}.   
Another approach was also developed, which enables us to treat directly the IR finite $S$-matrix \cite{Chung:1965zza, doi:10.1063/1.1664582, Kibble:1969ip, Kibble:1969ep, Kibble:1969kd, Kulish:1970ut}. It is called the dressed state formalism, which will be reviewed in the next subsection. 

Although the dressed state formalism was proposed many years ago, it has been recently reconsidered in the connection with the asymptotic symmetry (see, e.g., \cite{Mirbabayi:2016axw, Gabai:2016kuf, Kapec:2017tkm, Choi:2017bna, Choi:2017ylo, Carney:2018ygh, Neuenfeld:2018fdw}). It has been recognized that QED has an infinite number of symmetries associated with large gauge transformations \cite{He:2014cra, Campiglia:2015qka}. Thus, the conservation laws should constrain the $S$-matrix. On the other hand, scattering amplitudes vanish in the conventional approach because the sum of IR divergences at all orders produces the exponential suppression. 
It was pointed out in \cite{Kapec:2017tkm} that the vanishing of the amplitudes is consistent with the asymptotic symmetry of QED. Initial and final states used in the conventional approach generally belong to different sectors with respect to the asymptotic symmetry. Therefore, the amplitude between them should vanish since otherwise it breaks the conservation law. 
It was argued that we need dressed states in order to obtain non-vanishing amplitudes \cite{Kapec:2017tkm}. 

Motivated by these facts, we will investigate the dressed state formalism in this paper. 
In particular, we will revisit the gauge invariance in the formalism. 
We will argue that there is a problem on the gauge invariant condition in \cite{Kulish:1970ut}, and will resolve the problem.  In our method, dressed states are obtained just from the appropriate gauge invariant condition.
We will also discuss the $i\epsilon$ prescription for the dressed states. In addition, the relation between the dressed state formalism and the asymptotic symmetry will be considered.   
In order to explain our results more precisely, we first review the dressed state formalism in subsection~\ref{subsec:Smat}. 
We then present our results and the outline of the paper in subsection~\ref{subsec:summary}. 

\subsection{Review of a problem of $S$-matrix in QED and the dressed state formalism}\label{subsec:Smat}

$S$-matrix elements of scatterings in quantum field theories are defined by  
inner products of in-states and out-states: 
\begin{align}
\label{eq:Smat_elem}
S_{\alpha,\beta}=\outin,
\end{align}
where $\ket{\alpha}_{out}$ and $\ket{\beta}_{in}$ are eigenstates with energies $E_\alpha$ and $E_\beta$ of the Hamiltonian $H$ (which is assumed to be time-independent) such that they can be regarded as eigenstates $\ket{\alpha}_{0}$ and $\ket{\beta}_{0}$ with the same energies $E_\alpha$ and $E_\beta$ of a free Hamiltonian $H_0$ at $t\to \pm \infty$. 
More precisely, we should consider wave packets which are superpositions of eigenstates with a smooth function $g$ as follows:  
\begin{align}
\int d\beta\, g(\beta)\ket{\beta}_{in}. 
\end{align}
The condition of in-states and out-states are then given by   
\begin{align}
&\lim_{t_i \to -\infty}e^{-iH (t_i-t_s)}\int d\beta\, g_{in}(\beta)\ket{\beta}_{in}
=\lim_{t_i \to -\infty}e^{-iH_0 (t_i-t_s)}\int d\beta\, g_{in}(\beta)\ket{\beta}_{0},\\
&\lim_{t_f \to +\infty}e^{-iH (t_f-t_s)}\int d\alpha\, g_{out}(\alpha)\ket{\alpha}_{out}
=\lim_{t_f \to +\infty}e^{-iH_0 (t_f-t_s)}\int d\alpha\, g_{out}(\alpha)\ket{\alpha}_{0},
\end{align}
where we have introduced an arbitrary finite time $t_s$ at which the Schr\"odinger operators are defined. We can formally write the condition as 
\begin{align}
\label{eq:scat-states}
&\ket{\beta}_{in} = \lim_{t_i \to -\infty} \Omega(t_i) \ket{\beta}_{0}, \quad 
\ket{\alpha}_{out} = \lim_{t_f \to +\infty}  \Omega(t_f) \ket{\alpha}_{0},\\
\label{eq:moller}
&\Omega(t)\equiv U(t_s,t)U_0(t,t_s). 
\end{align}
Here, $U(t,t')$ and $U_0(t,t')$ are the full and free time-evolution operators respectively: 
\begin{align}
U(t,t')\equiv e^{-iH(t-t')}, \quad U_0(t,t')\equiv e^{-iH_0(t-t')}.
\end{align}
Using eq.\eqref{eq:scat-states}, the $S$-matrix element \eqref{eq:Smat_elem} can be written as 
\begin{align}
S_{\alpha,\beta}=\lim_{t_f\to \infty,\, t_i\to -\infty}\outfree{\alpha}\Omega(t_f)^\dagger\, \Omega(t_i)\ket{\beta}_{0}.
\end{align}
Since $\ket{\alpha}_{0}$ and $\ket{\beta}_{0}$ usually have the Fock state representation, we finally obtain the $S$-matrix operator on the Fock space $\mathcal{H}_{Fock}$ as 
\begin{align}
S=\lim_{t_f\to \infty,\, t_i\to -\infty}\Omega(t_f)^\dagger\, \Omega(t_i)
=\lim_{t_f\to \infty,\, t_i\to -\infty}
e^{i H_0(t_f-t_s)}e^{-iH(t_f-t_i)}e^{-i H_0(t_i-t_s)}.
\label{eq:S-mat_Fock}
\end{align}

For computations of the Fock space basis $S$-matrix \eqref{eq:S-mat_Fock}, it is convenient to play in the interaction picture. We divide the Hamiltonian as $H= H_0+V$, and define the interaction operator in the interaction picture as 
\begin{align}
V^I(t)\equiv U_0(t,t_s)^{-1}\, V\, U_0(t,t_s). 
\end{align}
Then the operator $\Omega(t)$ in \eqref{eq:moller} can be written as 
\begin{align}
\Omega(t)= \mathrm{T}\exp\left(-i\int^{t_s}_t  \!\!\!dt'\, V^I(t')\right),
\label{eq:Moller_int}
\end{align}
where the symbol $\mathrm{T}$ represents the time-ordered product. The $S$-matrix \eqref{eq:S-mat_Fock} can be represented as the Dyson series \cite{Dyson:1949ha}
\begin{align}
S= \mathrm{T}\exp\left(-i\int^{+\infty}_{-\infty} \!\!\!dt'\, V^I(t')\right).
\label{eq:Dyson_Smat}
\end{align}

The above is the standard treatment of the $S$-matrix in QFTs. However, this $S$-matrix on the Fock space is not well-defined in QED because of the infrared (IR) divergences. If we try to compute the $S$-matrix elements on the Fock space by the standard perturbation theory, we encounter the IR divergences. 

One way to address this problem is giving up the $S$-matrix as usually adopted in QFT textbooks such as \cite{Peskin:1995ev, Weinberg:1995mt}. It is argued that in any experiment for particle physics the detector has a minimum energy $E_d$ such that photons with energies less than $E_d$ cannot be detected, and therefore the measured cross-section is the sum of cross-sections for all events emitting undetectable soft photons \cite{Bloch:1937pw, Yennie:1961ad}. This inclusive method actually works and the measured cross-section is IR finite. 

Nevertheless, it is better that we have a well-defined $S$-matrix. Fortunately, there is a way to define an IR finite $S$ matrix. It is called the dressed state formalism \cite{Chung:1965zza, doi:10.1063/1.1664582, Kibble:1969ip, Kibble:1969ep, Kibble:1969kd, Kulish:1970ut}. 
IR divergences in the conventional approach originate from the assumption that the asymptotic scattering states can be regarded as free particle states at $t \sim \pm \infty$. Since photons are massless particles, \textit{i.e.}, the electromagnetism is a long-range interaction, we should take account of the interaction even in the asymptotic region. 
It means that we should modify the free time-evolution operator $U_0$ in \eqref{eq:moller} into another time-evolution operator $U_{as}$ which contains contributions of the long-range interaction in the asymptotic region. 
In fact, even for scatterings in quantum mechanics (not QFTs), in order to obtain the IR finite $S$-matrix in the Coulomb potential, we need such a modification \cite{Dollard1964}. 

In Faddeev and Kulish's paper \cite{Kulish:1970ut}, it was argued that the asymptotic dynamics of QED can be approximated by the following ``interacting'' Hamiltonian in the Schr\"odinger picture:
\begin{align}
\label{eq:asymptH}
H^s_{as}(t) = H^s_{0}+V_{as}^s(t) \quad \text{with}\quad V_{as}^s(t)=- \intx  A^s_\mu(\vx)j^\mu_{cl}(t,\vx),
\end{align}
where the superscript $s$ denotes that operators are in the Schr\"odinger picture, and $H^s_{0}$ is the usual free Hamiltonian for QED.  $j^\mu_{cl}(t,\vx)$ is a ``classical" current operator given by 
\begin{align}
\label{eq:cl_current}
j^\mu_{cl}(t,\vx)&= \sum e \intp \frac{p^\mu}{E_p}\, \delta^3(\vx -\vp t/E_p) \rho(\vp),
\\
\rho(\vp)&=b^\dagger(\vp)b(\vp)-d^{\dagger}(\vp)d(\vp),
\end{align}
where the sum in \eqref{eq:cl_current} runs over all charged particles, and we omit the label for simplicity. $b^\dagger$ (and $d^{\dagger}$) are creation operators of the charged particles (and antiparticles).\footnote{The creation and annihilation operators have labels for a spinor basis, if the particle is a fermion.} 
This current is ``classical" in the sense that it is a diagonal operator on the usual Fock space. 
Because of the explicit time-dependence of $j^\mu_{cl}$, the asymptotic Hamiltonian $H^s_{as}$ is time-dependent even in the Schr\"odinger picture. 

The $S$-matrix is then given by 
\begin{align}
S=\lim_{t_f\to \infty,\, t_i\to -\infty}\Omega_{as}(t_f)^\dagger\, \Omega_{as}(t_i)
=\lim_{t_f\to \infty,\, t_i\to -\infty}
U_{as}^\dagger(t_f,t_s)e^{-iH(t_f-t_i)}U_{as}(t_i,t_s),
\label{eq:FK_Smat}
\end{align}
where 
$\Omega_{as}(t)$ is obtained by replacing $U_0$ in \eqref{eq:moller} into $U_{as}$ with
\begin{align}
U_{as}(t,t_s)\equiv \mathrm{T}\exp\left(-i\int^t_{t_s}  \!\!\!dt'\, H^s_{as}(t')\right).
\label{eq:Uas}
\end{align}  

We can proceed further by computing this asymptotic time-evolution operator \eqref{eq:Uas}. Similar to the derivation of \eqref{eq:Moller_int}, one can find (see \cite{Kulish:1970ut} for the derivation) that $U_{as}(t,t_s)$ is given by 
\begin{align}
U_{as}(t,t_s)=U_0(t,t_s) \, \mathrm{T}\exp\left(-i\int^t_{t_s}  \!\!\!dt'\, V^I_{as}(t')\right),
\end{align}
where 
\begin{align}
V^I_{as}(t)\equiv U_0(t,t_s)^{-1}\, V^s_{as}(t)\, U_0(t,t_s). 
\end{align}
Furthermore, since the commutator $[V^I_{as}(t_1),V^I_{as}(t_2)]$ 
commutes with $V^I_{as}(t)$ for any $t$, we obtain 
\begin{align}
U_{as}(t,t_s)=U_0(t,t_s)\, e^{-i\int^t_{t_s}  \!\!\!dt'\, V^I_{as}(t')} e^{-\frac{1}{2} \int^t_{t_s} \!\!\!dt_1 \int^{t_1}_{t_s} \!\!\!dt_2 \,[V^I_{as}(t_1),V^I_{as}(t_2)]},
\label{eq:Uas_V}
\end{align}
and by performing the $t$-integral, we have
\begin{align}
-i\int^t_{t_s}  \!\!\!dt'\, V^I_{as}(t')=R(t)-R(t_s)
\label{eq:R(t_s)}
\end{align}
with 
\begin{align}
\label{eq:FK_R1}
R(t)\equiv \sum e \intp \rho(\vp)\intkw \frac{p^\mu}{p \cdot k}\left[
a_\mu(\vk)e^{i \frac{p \cdot k}{E_p}t}
-a_\mu^{\dagger}(\vk)e^{-i \frac{p\cdot k}{E_p}t}
\right],
\end{align}
where $a_\mu (\vk)$ are annihilation operators of photons  $[k^\mu=(\omega, \vk),\,\omega=|\vk|]$. 
The exponent including the commutator $[V^I_{as}(t_1),V^I_{as}(t_2)]$ in \eqref{eq:Uas_V} is a classical operator in the same sense as $j^\mu_{cl}(t,\vx)$, and we represent it as $i\Phi(t,t_s)$ where
\begin{align}
\label{eq:FK_phi1}
\Phi(t,t_s) &\equiv \frac{i}{2} \int^t_{t_s} \!\!\!dt_1 \int^{t_1}_{t_s} \!\!\!dt_2 \,[V^I_{as}(t_1),V^I_{as}(t_2)].
\end{align}
In \cite{Kulish:1970ut}, $R(t_s)$ in \eqref{eq:R(t_s)} was deleted by a requirement for an initial condition.
Permitting this,\footnote{We will see that we do not need to worry about this requirement in our approach.} the $S$-matrix \eqref{eq:FK_Smat} becomes 
\begin{align}
S=\lim_{t_f\to \infty,\, t_i\to -\infty} 
e^{-R(t_f)} e^{-i \Phi(t_f,t_s)}\,\left[ \mathrm{T}\exp\left(-i\int^{t_f}_{t_i} \!\!\!dt'\, V^I(t')\right) \right] e^{R(t_i)} e^{i \Phi(t_i,t_s)}.
\end{align}
As a result, this $S$-matrix differs from usual Dyson's one \eqref{eq:Dyson_Smat} only in the dressing factors $e^R$ and $e^{i\Phi}$. Thus, if we formally introduce a dressed Hilbert space $\mathcal{H}_{FK}$ as 
\begin{align}
\mathcal{H}_{FK} =\lim_{t\to -\infty} e^{R(t)} e^{i\Phi(t,t_s)} \mathcal{H}_{Fock},
\label{eq:FK_Hilbert}
\end{align}
the $S$-matrix on $\mathcal{H}_{FK}$ is given by usual one \eqref{eq:Dyson_Smat}.\footnote{Even on $\mathcal{H}_{FK}$, the notion of particles for charged fields is still valid because $j^\mu_{cl}(t,\vx)$ in the dressing factor $e^{R(t)}$ is a diagonal operator on the Fock space. However, the standard interpretation of photons on the Fock space seems to be lost because the dressing factor excites an infinite number of photons. As we will see in subsec.~\ref{sec:IRfinite}, the energy  of the excited photons by the dressing factor is soft in the limit $t\to \pm \infty$. Hence, the particle notion for hard photons may be valid.} 
Actually, the factors play a similar role as summing the contributions of soft photons, and the $S$-matrix on $\mathcal{H}_{FK}$ is known to be IR finite \cite{Chung:1965zza},
if we impose the physical state condition. 
In subsec.~\ref{sec:IRfinite}, we will comment on a subtlety of the proof of IR finiteness in \cite{Chung:1965zza}.

\subsection{Our method and the differences from Faddeev and Kulish's}\label{subsec:summary}

Not all of the states in $\mathcal{H}_{FK}$ are physical, and thus we have to restrict $\mathcal{H}_{FK}$ to the subspace by imposing a gauge invariant condition. 
However, the treatment for the gauge invariance in \cite{Kulish:1970ut}  seems inappropriate. The \textit{free} Gupta-Bleuler condition was imposed on $\mathcal{H}_{FK}$ as the physical state condition, \textit{i.e.}, physical states, $\ket{\psi} \in \mathcal{H}_{FK}$, were required to satisfy 
\begin{align}
k^\mu a_\mu (\vk) \ket{\psi} =0 \quad \text{for any $\vk$} .
\label{eq:freeGB}
\end{align}
In \cite{Kulish:1970ut}, to satisfy \eqref{eq:freeGB},  the dressing operator $R$ in \eqref{eq:FK_R1} was modified by introducing a null vector $c^\mu(\vk)$ satisfying $k_\mu c^\mu=1$. 
More concretely, the dressing operator was altered by shifting the coefficient $\frac{p^\mu}{p \cdot k}$ in \eqref{eq:FK_R1} to $\frac{p^\mu}{p \cdot k}-c^\mu$.

We will see that the artificial vector $c_\mu$ is not needed for an appropriate gauge invariant condition. 
Our claim is that contributions of long-range interactions should be incorporated into the gauge invariant condition, as dressed states are obtained by taking account of such an interaction. 
The free Gupta-Bleuler condition \eqref{eq:freeGB} is not adequate for dressed states. 
In section~\ref{sec:Gauge_inv}, we will present the appropriate condition.

Furthermore, we will show that the dressed Hilbert space can be obtained just by requiring the gauge invariant condition.  In our approach, it turns out that we do not need to solve the dynamics of the asymptotic Hamiltonian $H_{as}$ as we reviewed in subsec~\ref{subsec:Smat}. In fact, although the Dyson's $S$-matrix \eqref{eq:Dyson_Smat} is not a good operator on the usual Fock space $\mathcal{H}_{Fock}$, it is well-defined on the dressed space $\mathcal{H}_{FK}$.\footnote{In this paper, we do not take care of problems at ultraviolet regions. We assume that they can be resolved by a standard renormalization procedure.}  
The asymptotic Hamiltonian $H_{as}$ is just an approach to derive the dressing factor $e^{R(t)}$. We think that the gauge invariant condition is a simpler approach to obtain the factor, and the interpretation is clear. The condition essentially just says that if there is a charged particle, there should exist electromagnetic fields around it by Gauss's law. The fields around the charge indeed make up the dress. 

In section \ref{sec:meaning}, as a support of this interpretation and also another justification that we do not have to introduce $c_\mu$, we will discuss the meaning of the original dressing operator $R(t)$ in eq.~\eqref{eq:FK_R1}. As shown in \cite{Bagan:1999jf}, the dressing factor for a charged particle with momentum $\vp$ corresponds to the Li\'enard-Wiechert potential for the uniformly moving charge with momentum $\vp$. 
We will reconfirm this fact especially taking care of the $i\epsilon$ prescription.

Besides, our method allows a variety of dresses, and $\mathcal{H}_{FK}$ given by \eqref{eq:FK_Hilbert} is just one of them. We will see that in our gauge invariant condition, the physical Hilbert space $\mathcal{H}_{as}$ on which Dyson's $S$-matrix \eqref{eq:Dyson_Smat} acts takes the form 
\begin{align}
\mathcal{H}_{as}= e^{R_{as}} \mathcal{H}_{free},
\label{eq:general_dressed_states}
\end{align}
where $e^{R_{as}}$ is a dressing factor, and $\mathcal{H}_{free}$ is a subspace of the Fock space $\mathcal{H}_{Fock}$ such that the free Gupta-Bleuler condition are satisfied $(k^\mu a_\mu (\vk) \ket{\psi} =0,\, \ket{\psi} \in \mathcal{H}_{free})$.
The operator $R_{as}$ can be $R(t)+i\Phi(t,t_s)$, but not necessary.
We will discuss the relation between the ambiguity of dressing and the asymptotic symmetry of QED in subsec~\ref{sec:asympt}.

We conclude that infrared divergences of the $S$-matrix in the usual perturbative approach for QED are caused by the usage of the inappropriate asymptotic states.  
Although the asymptotic states satisfying the free Gupta-Bleuler condition may be used at the tree level, they should not be used at loop levels.
If we instead use the correct gauge-invariant states at the same order, we can avoid IR divergences. 
 
\section{Necessity of dresses}\label{sec:Gauge_inv}
We show that states with charged particles must include photons even in the interaction picture to satisfy the gauge invariance. In order to impose the physical state condition in a systematic way, we use the BRST formalism. 

\subsection{Lagrangian and Hamiltonian in covariant gauge} 
In the BRST formalism with the Feynman gauge, the Lagrangian of QED is given by
\begin{align}
\mathcal{L}_{QED}=\mathcal{L}_{EM}+\mathcal{L}_{matter}+\mathcal{L}_{GF}+\mathcal{L}_{FP},
\label{full-L}
\end{align}
where
\begin{align}
&\mathcal{L}_{EM}=-\frac{1}{4}F_{\mu\nu}F^{\mu\nu}\,,\quad  
\mathcal{L}_{GF}=-\frac{1}{2}\bigl(\partial^{\mu}A_{\mu}\bigr)^{2}\ ,\quad
\mathcal{L}_{FP}=i\,\partial^{\mu}\bar{c}\,\partial_{\mu}c,
\end{align}
and the metric signature in this paper is $(-,+,+,+)$.
Here we have integrated out the Nakanishi-Lautrap field. The ghost field $c$ can also be integrated out since it is completely decoupled in QED. However, we keep it to obtain the BRST charge.
$\mathcal{L}_{matter}$ is the Lagrangian of charged matter fields. In this paper, matter fields can be any massive complex scalars and fermions without derivative self-interactions, and we do not write down the explicit form of the Lagrangian. They are coupled to the gauge field so that the EoM of the gauge field is given by 
\begin{align}
\Box A^\mu=-j^\mu,
\end{align}
where $j^\mu$ is the matter current derived from $\mathcal{L}_{matter}$. For example, if we have a charged  scalar $\phi$ with charge $e$ such as 
\begin{align}
\label{eq:ex_scalar_Lag}
\mathcal{L}_{\phi}=-D_{\mu}\bar{\phi}D^{\mu}\phi-m^2 \bar{\phi}\phi - V_{\phi}(|\phi|),
\end{align}
with $D_\mu \phi = \partial_\mu \phi - i e A_\mu \phi$, the matter current is given by 
\begin{align}
j^\mu=ie\bigl(D^{\mu}\bar{\phi}(x)\,\phi(x)-\bar{\phi}(x)\,D^{\mu}\phi(x)\bigr).
\end{align}

We now consider the Hamiltonian. 
We represent the conjugate momentum fields of $A^\mu, c, \bar{c}$ by $\Pi_\mu, \pc, \bpc$ which are defined from the Lagrangian \eqref{full-L} as 
\begin{align}
\Pi_0=-\partial_\mu A^\mu, \quad
\Pi_i=F_{0i}, \quad
\pc=-i \partial_0 \bar{c}, \quad 
\bpc=i \partial_0 c.
\end{align}
The Hamiltonian is then given by free part $H_0$ and the other interacting part $V$: 
\begin{align}
H=H_0+ V\,,
\label{eq:totH}
\end{align}
where  
\begin{align}
&H_0= H_{EM}+ H_{matter}+H_{ghost},
\end{align}
with\footnote{\label{foot:ham}
	$H_{EM}$ given by \eqref{eq:Hem} is different from the Hamiltonian obtained in a canonical way from Lagrangian \eqref{full-L}  by a total derivative term. We have eliminated the boundary term, and then this $H_{EM}$ commutes with the BRST charge without a boundary term. 
	This difference is not important except in sec~\ref{sec:asympt}.
}   
\begin{align}
\label{eq:Hem}
&H_{EM}= \intx \left[\frac12 \Pi_\mu \Pi^\mu +(\partial_i\Pi_0)   A^i +(\partial_i \Pi_i)  A^0 +\frac14 F_{ij} F^{ij}\right]\,,\\
&H_{ghost} = i \intx (\bpc \pc-\partial_i \bar{c} \partial_i c),
\end{align}
and $H_{matter}$ is the free part of the Hamiltonian of matter fields.

\subsection{Physical states in the Schr\"odinger picture} 
We now quantize the system by imposing the canonical (anti)commutation relations. We put the superscript $s$ to represent that an operator is in the Schr\"odinger picture like $A_\mu^s$. The equal-time (anti)commutation relations for the gauge fields and the ghost fields are given by\footnote{The ghost fields $c$ and $\bar{c}$ are not related by the Hermitian conjugation. They are Grassmann-odd Hermitian operators $c^{s\dagger}=c^s,\, \bar{c}^{s\dagger}=\bar{c}^s$ and their conjugate momenta are anti-Hermitian $\pc^{s\dagger}=-\pc^s,\, \bpc^{s\dagger}=-\bpc^s$. The Lagrangian and the Hamiltonian are real under this Hermiticity.} 
\begin{align}
\label{eq:canon_com}
[A_\mu^s(\vx), \Pi_\nu^s(\vy)] &= i \eta_{\mu\nu} \delta^3(\vx-\vy)\,,\quad 
\{c^s(\vx), \pc^s(\vy)\}=\{\bar{c}^s(\vx), \bpc^s(\vy)\}=i \delta^3(\vx-\vy).
\end{align}

The obtained Hilbert space is too large, and the physical Hilbert space $\mathcal{H}_{phys}$ is given by the BRST cohomology. In the Schr\"odinger picture, the BRST operator is expressed as 
\begin{align}
\label{eq:BRST_pos}
Q_{BRST}^s=-\intx \left[i \bpc^s \Pi^{0s}-\partial_i c^s \Pi^{is}+ c^s j^{0s}
\right].
\end{align}
Note that, since the BRST charge acts on charged matter fields, it has a term containing the matter current $j^{\mu s}$. For example, the BRST operator acts on the charged scalar in \eqref{eq:ex_scalar_Lag} as 
\begin{align}
[Q_{BRST}^s, \phi^s]=e c^s \phi^s, 
\end{align}
which is the BRST transformation for the charged scalar. 
This BRST charge commutes with the total Hamiltonian and the matter current
\begin{align}
[Q_{BRST}^s, H^s]=0\,, \quad [Q_{BRST}^s, j^{\mu s}(\vx)]=0.
\end{align}

It is convenient to move to the momentum representation because the ghost fields are free. If we write 
\begin{align}
c^s(\vx)&= \intkw \left[c(\vk) e^{-i\omega t_s +i \kx} +c^\dagger(\vk) e^{i\omega t_s -i \kx}
\right], 
\\
\bar{c}^s(\vx)&= \intkw \left[\bar{c}(\vk) e^{-i\omega t_s +i \kx} +\bar{c}^\dagger(\vk) e^{i\omega t_s -i \kx}
\right], 
\\
\pc^s(\vx)&= -\intk \frac{1}{2}\left[\bar{c}(\vk) e^{-i\omega t_s +i \kx} -\bar{c}^\dagger(\vk) e^{i\omega t_s -i \kx}
\right], 
\\
\bpc^s(\vx)&= \intk \frac{1}{2}\left[c(\vk) e^{-i\omega t_s +i \kx} -c^\dagger(\vk) e^{i\omega t_s -i \kx}
\right],
\end{align}
eq.~\eqref{eq:canon_com} leads to  
\begin{align}
\{c(\vk), \bar{c}^\dagger(\vkp)\} =i (2\omega)(2\pi)^3  \delta^3(\vk-\vkp),\quad 
\{\bar{c}(\vk), c^\dagger(\vkp)\} =-i(2\omega)(2\pi)^3  \delta^3(\vk-\vkp).
\end{align}
We also introduce the ladder operators of photons as\footnote{This is always possible because this is just a change of canonical variables at a time $t_s$.} 
\begin{align}
&A_\mu^s(\vx)=\intkw \left[a_\mu(\vk)e^{-i\omega t_s +i \kx} +a_\mu^{\dagger}(\vk) e^{i\omega t_s -i \kx}\right],
\\
&\Pi_0^s(\vx)=-i \intkw \left[k^\mu a_\mu(\vk)e^{-i\omega t_s +i \kx} -k^\mu a_\mu^{\dagger}(\vk) e^{i\omega t_s -i \kx}\right],
\\
&\Pi_i^s(\vx)=-i \intkw \left[(k_i a_0(\vk)+\omega a_i(\vk))e^{-i\omega t_s +i \kx} -(k_i a_0^{\dagger}(\vk)+\omega a_i^{\dagger}(\vk))e^{i\omega t_s -i \kx}\right],
\end{align}
with
\begin{align}
[a_\mu(\vk),a_\nu^{\dagger}(\vkp)]=(2\omega)(2\pi)^3  \delta^3(\vk-\vkp)\eta_{\mu\nu}.
\end{align}

The BRST operator \eqref{eq:BRST_pos} is then written as 
\begin{align}
Q_{BRST}^s=-\intkw \left[
c(\vk) \{k^\mu a_\mu^{\dagger}(\vk) +e^{-i\omega t_s}\tilde{j}^{0s}(-\vk)\}
+c^\dagger(\vk)  \{k^\mu a_\mu(\vk) +e^{i\omega t_s}\tilde{j}^{0s}(\vk)\}
\right],
\end{align}
where $\tilde{j}^{0s}(\vk)$ is the Fourier transformation of $j^{0s}(\vx)$ defined as 
\begin{align}
\tilde{j}_\mu^{s}(\vk)=\intx \,e^{-i \kx} j_\mu^s(\vx).
\end{align}

Since the ghost fields are decoupled, we can always restrict the ghost-sector of physical states to the ghost-vacuum annihilated by $c(\vk)$ and $\bar{c}(\vk)$. Therefore, the physical state condition $Q^s_{BRST} \ket{\psi}=0$ becomes 
\begin{align}
\label{eq:phys_cond_schr}
\left[ k^\mu a_\mu(\vk)+e^{i\omega t_s}\tilde{j}^{0s}(\vk)\right]\ket{\psi}=0 \quad \text{for all $\vk$} \,.
\end{align}
This is the physical state condition in the Schr\"odinger picture. Note that this condition is different from the usual Gupta-Bleuler condition \eqref{eq:freeGB}. This fact holds in the interaction picture as we will see in next subsection.

\subsection{Gauge invariant asymptotic states}\label{sec:asymptostates}
We now move to the interaction picture such that the $S$-matrix is given by the usual one \eqref{eq:Dyson_Smat}. The asymptotic state $\ket{\beta}_{0}$ which is acted on the $S$-matrix is related to the in-states in the Schr\"odinger picture, $\ket{\beta}_{in}$,  as \eqref{eq:scat-states}. Since $\ket{\beta}_{in}$ is a physical state in the Schr\"odinger picture, it satisfies $Q^s_{BRST}\ket{\beta}_{in}=0$. 
Thus, from eq.~\eqref{eq:scat-states}, $\ket{\beta}_{0}$ should satisfy 
\begin{align}
0=Q^s_{BRST}\ket{\beta}_{in}= \lim_{t_i \to -\infty} Q^s_{BRST} \Omega(t_i) \ket{\beta}_{0}.
\end{align}
Since $Q^s_{BRST}$ commutes with the exact Hamiltonian $H^s$, we have 
\begin{align}
Q^s_{BRST} \Omega(t)= U(t_s,t) Q^s_{BRST} U_0(t,t_s)=\Omega(t) Q^I_{BRST}(t),
\end{align}
where $Q^I_{BRST}(t)$ is the BRST operator in the interaction picture: 
\begin{align}
Q^I_{BRST}(t) &\equiv U_0(t,t_s)^{-1}\,Q^s_{BRST}\,U_0(t,t_s)\nn
&=-\intkw \left[
c(\vk) \{k^\mu a_\mu^{\dagger}(\vk) +e^{-i\omega t}\tilde{j}^{0I}(t,-\vk)\}
+c^\dagger(\vk)  \{k^\mu a_\mu(\vk) +e^{i\omega t}\tilde{j}^{0I}(t,\vk)\}
\right]
.
\end{align}
Therefore, $\ket{\beta}_{0}$ satisfies 
\begin{align}
\lim_{t_i \to -\infty} Q^I_{BRST}(t_i) \ket{\beta}_{0}=0.
\end{align}
By restricting the ghost-sector to the ghost-vacuum, this condition becomes 
\begin{align}
\lim_{t_i \to -\infty}\left[ k^\mu a_\mu(\vk)+e^{i\omega t_i}\tilde{j}^{0I}(t_i,\vk)\right]\ket{\beta}_{0}=0. 
\label{eq:int_phys_cond}
\end{align}
It means that states satisfying the free Gupta-Bleuler condition $k^\mu a_\mu (\vk) \ket{\psi} =0$ are generally not the physical asymptotic states. Thus, the charged 1-particle states in the standard Fock space, such as $b^\dagger(\vp)\ket{0}$, cannot be the asymptotic physical states. 

We will show below that the states satisfying the condition \eqref{eq:int_phys_cond} are dressed states. 
In fact, if there is an anti-Hermitian operator $\tilde{R}(t)$ such that 
\begin{align}
[k^\mu a_\mu (\vk),\tilde{R}(t)]=-e^{i\omega t} \tilde{j}^{0I}(t,\vk), \quad [\tilde{j}^{0I}(t,\vk),\tilde{R}(t)]=0,
\label{eq:dress_cond_I}
\end{align}
then states in $e^{\tilde{R}(t)}\mathcal{H}_{free}$ are annihilated by $k^\mu a_\mu(\vk)+e^{i\omega t}\tilde{j}^{0I}(t,\vk)$ where $\mathcal{H}_{free}$ is a subspace of $\mathcal{H}_{Fock}$ satisfying the free Gupta-Bleuler condition. Thus, the Hilbert space satisfying \eqref{eq:int_phys_cond} is given by 
\begin{align}
\lim_{t_i \to -\infty}e^{\tilde{R}(t_i)}\mathcal{H}_{free}.
\end{align}
There are various choices of the dressing operator satisfying \eqref{eq:dress_cond_I}. One example is 
\begin{align}
\label{eq:footnote_R}
\tilde{R}(t)=\intkw \frac{1}{2\omega^2}\left[
e^{i\omega t}\tilde{j}^{0I}(t,\vk) \tilde{k}^\mu a^\dagger_\mu(\vk)-e^{-i\omega t}\tilde{j}^{0I}(t,-\vk) \tilde{k}^\mu a_\mu(\vk)
\right],
\end{align}
where $\tilde{k}^\mu=(\omega,-\vk)$. 

Although one may use such a dressing operator $\tilde{R}(t)$, we can simplify it by recognizing that the current operator $j^{0 I}$ can be approximated in the asymptotic regions $(t \sim \pm\infty)$ by the classical current operator $j^{0}_{cl}$ given by \eqref{eq:cl_current}. It is convenient to use the hyperbolic coordinates $(\tau,\rho)$ to look at the asymptotic behaviors of massive charged fields as follows \cite{Campiglia:2015qka}: 
\begin{align}
t=\sqrt{1+\rho^2}\,\tau, \quad \vx=|\tau|\rho \hat{x}.
\end{align}
Then we can straightforwardly obtain (see appendix C in  \cite{Hirai:2018ijc} for details)\footnote{Other components of the current also satisfy similar equations: 
\begin{align}
\lim_{\tau\to \pm\infty}j^{i I}_{free}(t,\vx)=\lim_{\tau\to \pm \infty}j^{i}_{cl}(t,\vx).
\end{align}
Here, the subscript $free$ means that the current is that of the free theory.} 
\begin{align}
\lim_{\tau\to \pm\infty}j^{0 I}(t,\vx)=\lim_{\tau\to \pm \infty}j^{0}_{cl}(t,\vx).
\end{align}
Therefore, we can rewrite the condition \eqref{eq:int_phys_cond} as 
\begin{align}
\label{eq:as_phys_cond} 
\lim_{t_i \to -\infty}\left[ k^\mu a_\mu(\vk)+e^{i\omega t_i}\tilde{j}^{0}_{cl}(t_i,\vk)\right]\ket{\beta}_{0}=0. 
\end{align}
For later convenience, we represent the operator in \eqref{eq:as_phys_cond} by $\hat{G}(t,\vk)$ as 
\begin{align}
\hat{G}(t,\vk)\equiv k^\mu a_\mu(\vk)+e^{i\omega t}\tilde{j}^{0}_{cl}(t,\vk).
\end{align}

Noting that the momentum representation of the classical current operator is given by 
\begin{align}
\tilde{j}^0_{cl}(t,\vk)&= \sum e \intp e^{-i\frac{\vp\cdot\vk}{E_p}t} \rho(\vp),
\end{align}
and a trivial equation $e^{-i \frac{p \cdot k}{E_p}t}=e^{i \omega t}e^{-i\frac{\vp\cdot\vk}{E_p}t}$, 
we can easily confirm that the Faddeev-Kulish dressing operator $R(t)$ in \eqref{eq:FK_R1} satisfies
\begin{align}
\hat{G}(t,\vk)e^{R(t)}=e^{R(t)}\,k^\mu a_\mu(\vk). 
\end{align}
Thus, an asymptotic physical Hilbert space satisfying \eqref{eq:as_phys_cond} is given by 
\begin{align}
\lim_{t_i \to -\infty}e^{R(t_i)}\,\mathcal{H}_{free}.
\end{align}

Since the phase operator $\Phi$ in \eqref{eq:FK_phi1} commutes with $\hat{G}(t,\vk)$ and $R(t)$, $\Phi$ is not relevant for the gauge-invariance \eqref{eq:as_phys_cond}.  Therefore, the Faddeev-Kulish dressed space $\mathcal{H}_{FK}$ in \eqref{eq:FK_Hilbert} is gauge invariant without introducing a vector $c_\mu$, if we restrict $\mathcal{H}_{Fock}$ to the subspace $\mathcal{H}_{free}$.

Besides of the phase operator, there are other choices of the dressing operator $R_{as}(t)$ satisfying
\begin{align}
\hat{G}(t,\vk)e^{R_{as}(t)}=e^{R_{as}(t)}\,k^\mu a_\mu(\vk). 
\end{align}
One example of $R_{as}(t)$ other than the Faddeev-Kulish dressing operator \eqref{eq:FK_R1} is obtained by replacing $\tilde{j}^{0I}$ with $\tilde{j}^{0}_{cl}$ in \eqref{eq:footnote_R}. 
Then we can define another asymptotic physical Hilbert space:
\begin{align}
\lim_{t_i \to -\infty}e^{R_{as}(t_i)}\,\mathcal{H}_{free},
\end{align}
which is a solution of the gauge invariant condition \eqref{eq:as_phys_cond}. 
Although the question that what types of dressing cancel the IR divergences in the $S$-matrix is beyond the scope of this paper, we will discuss in subsection~\ref{sec:asympt} that 
the existence of many choices is natural from the point of view of asymptotic symmetry.

\section{Interpretation of the Faddeev-Kulish dresses}\label{sec:meaning}

It is shown in \cite{Bagan:1999jf} that the Faddeev-Kulish  dressing factor for a charged particle with momentum $p^\mu$ corresponds to the classical Li\'enard-Wiechert potential around the particle. This fact supports our statement that Gauss's law require the dressing factor. 
In this section, we will reconfirm this fact taking care of the $i\epsilon$ prescription, and see that we should use different prescriptions for initial and final states, which might be useful for the explicit computation of scattering amplitudes.

\subsection{Coulomb potential by point charges in the asymptotic region}
Here, we will recall the expression of the electromagnetic potential created by a charged point particle with momentum $p^\mu$. 
The classical equation of motion for the gauge field in the Lorenz gauge is given by 
\ba{
	\square A _ { \mu } ( x )=-j_{\mu}(x)\ ,\ j_{\mu}(x)=e\int^{\infty}_{-\infty}\!\!d\tau \frac{dy_{\mu}(\tau)}{d\tau}\delta^{4}(x-y(\tau)),\label{cleom}
}
where $y_{\mu}(\tau)=\frac{p_{\mu}}{m}\tau=\frac{p_{\mu}}{E_{p}}t$ is the trajectory of the charged particle, which is supposed to pass through the origin at $t=0$. The position at $t=0$ is not relevant when we consider the asymptotic region.\footnote{
 However, the position at $t=0$ can contribute to subleading orders, and it was shown in \cite{Hamada:2018cjj} that the position is important for the subleading memory effect.}
By using the retarded Green's function for the Klein-Gordon equation,
\ba{
	G_{ret}(x)=-\int \frac{d^{4}k}{(2\pi)^4}\frac{1}{(k^{0}-\omega+i\epsilon)(k^{0}+\omega+i\epsilon)}e^{ik\cdot x},
}
the general solutions of \eqref{cleom} are given by
\ba{
	A _ { \mu } ( x )&=A^{in}_{\mu}(x)+\int\! d^{4}x' G_{ret}(x-x')j_{\mu}(x')\nonumber\\
	&=A^{in}_{\mu}(x)+ie\frac{p_{\mu}}{E_{p}}\int\! \frac{d^{3}k}{(2\pi)^3}\int_{-\infty}^{t}\!\!\!dt' \frac{1}{2 \omega}\Big{(}e^{-i\omega \left(t-t'\right)}-e^{i\omega \left(t-t'\right)}\Big{)}e^{-\epsilon\left(t-t'\right)}e^{i\vec{k}\cdot \left(\vec{x}-\frac{\vec{p}}{E_{p}}t'\right)}\nonumber\\
	&=A^{in}_{\mu}(x)-e\int\! \frac{d^{3}k}{(2\pi)^{3} (2\omega)}\Bigl[ \frac{p_{\mu}}{p\cdot k+i\epsilon}e^{i\vec{k}\cdot \left(\vec{x}-\frac{\vec{p}}{E_{p}}t\right)}+\frac{p_{\mu}}{p\cdot k-i\epsilon}e^{-i\vec{k}\cdot \left(\vec{x}-\frac{\vec{p}}{E_{p}}t\right)}\Bigr] \label{Coulomb}.
}
where $A^{in}_{\mu}(x)$ is the incoming free wave, which is specified at $t\rightarrow -\infty$, and the second term is the Li\'enard-Wiechert potential created by the particle with momentum $p^\mu$ and charge $e$. 
We represent this second term by $A_\mu^{ret}(x;\vp)$ as
\begin{align}
A_\mu^{ret}(x;\vp) \equiv -e\int\! \frac{d^{3}k}{(2\pi)^{3} (2\omega)}\Bigl[ \frac{p_{\mu}}{p\cdot k+i\epsilon}e^{i\vec{k}\cdot \left(\vec{x}-\frac{\vec{p}}{E_{p}}t\right)}+\frac{p_{\mu}}{p\cdot k-i\epsilon}e^{-i\vec{k}\cdot \left(\vec{x}-\frac{\vec{p}}{E_{p}}t\right)}\Bigr].
\label{eq:ini_Coulomb}
\end{align}

\subsection{Coulomb potential from dressed states with $i\epsilon$ prescription}

Let's consider a dressed state of a single incoming electron with momentum $p^\mu$ defined by
\ba{
	|\!|p(t)\rangle\!\rangle\equiv e^{R_{in}(t)}b^{\dagger}(\vec{p})\ket{0},
	\label{eq:ini_1p}
}
where $R_{in}(t)$ is an operator dressing the incoming single particle state. 
The gauge field in the interaction picture can be written as
\be
A^I_{ \mu } ( x ) = \int\! \frac { d ^ { 3 } k } { ( 2 \pi ) ^ { 3 } (2 \omega)} \left( a _ { \mu } ( \vec { k } ) e ^ { i k\cdot x } + a _ { \mu } ^ { \dagger } ( \vec { k } ) e ^ { - i k\cdot x } \right).
\ee
Then we demand that its expectation value for the above dressed state\footnote{More precisely, we should use a wave-packet since the state \eqref{eq:ini_1p} is not normalized.} match the classical gauge field \eqref {eq:ini_Coulomb} created by a charged point particle with momentum $p^\mu$ as 
\begin{align}
    &\langle\!\langle p(t)|\!| A^I_{\mu}(x) |\!|p(t)\rangle\!\rangle
	\nn
	&=-e\int\!\frac{d^{3}k}{(2\pi)^{3} (2\omega)}\Big{(} \frac{p_{\mu}}{p\cdot k+i\epsilon}e^{i\vec{k}\cdot \left(\vec{x}-\frac{\vec{p}}{E_{p}}t\right)}+\frac{p_{\mu}}{p\cdot k-i\epsilon}e^{-i\vec{k}\cdot \left(\vec{x}-\frac{\vec{p}}{E_{p}}t\right)}\Big{)}
	\langle\!\langle p(t)|\!|p(t)\rangle\!\rangle.
	\label{consistencytocl}
\end{align}
We can easily check that the following dressing operator satisfies the above condition, 
\ba{
	R_{in}(t)=e\int\! \frac { d ^ { 3 } p } { ( 2 \pi ) ^ { 3 } (2 E_{p}) }\rho(\vec{p})\int\! \frac { d ^ { 3 } k } { ( 2 \pi ) ^ { 3 } (2 \omega) }\Big{(}\frac{p^{\mu}}{p\cdot k-i\epsilon}a_{\mu}(\vec{k})e^{i\frac{p\cdot k}{E_{p}}t}-\frac{p^{\mu}}{p\cdot k+i\epsilon}a^{\dagger}_{\mu}(\vec{k})e^{-i\frac{p\cdot k}{E_{p}}t}\Big{)}.
	\label{eq:ini_dress}
}
This operator matches the dressing operator \eqref{eq:FK_R1} up to  the $i\epsilon$ insertion. How to insert $i\epsilon$ in the dressing operator is determined by how the initial condition of gauge fields is specified. 
Thus, the dressed states stand for the states of (anti-)electrons surrounded by relativistic Coulomb fields created by themselves. 
We have considered a single charged particle state \eqref{eq:ini_1p}. The generalization to multi-particle states is trivial, and the expectation value of $A_\mu$ is given by the  superposition of the Coulomb field created by each particle. 
In other words, in the dressed state, the charged particles are properly dressed by electromagnetic fields in the asymptotic region where the particles have nearly constant velocities. This result is natural since our dressed states are obtained by solving the BRST (gauge invariant) condition without ignoring the interaction in the asymptotic regions.
We also would like to comment that this expectation value changes if we modify the dressing operator by introducing a vector $c_\mu$ as in \cite{Kulish:1970ut}. This is another reason to think that such a modification is unnatural. 
Note also that $R_{in}$ is anti-Hermitian ($R_{in}^\dagger=-R_{in}$). Thus, the dressing factor $e^{-R_{in}}$ is unitary.\footnote{
	If we write $e^{-R_{in}}$ in the normal ordering, the normalization factor has an IR divergence if we set $\epsilon=0$. Thus, it is often said (see, e.g., \cite{Kulish:1970ut}) that the dressing factor is not a unitary operator on the Fock space in a rigorous sense. However, it does not matter if we keep $\epsilon$ nonzero. After computing IR finite physical quantities, we can take $\epsilon$ to 0.} 

Similarly, we can fix the $i\epsilon$ prescription for the dressing operator $R_{out}(t)$ for outgoing states. 
We consider a dressed outgoing state 
\begin{align}
	~_{out} \langle\!\langle p(t)|\!|\equiv
	\bra{0} b(\vp) e^{-R_{out}(t)},
\end{align}
and require that the expectation value of $A^I_\mu(x)$ agree with the advanced potential for the point particle, 
which is given by 
\begin{align}
A^{adv}_\mu(x;\vp)=-e\int\! \frac{d^{3}k}{(2\pi)^{3} (2\omega)}\Bigl[ \frac{p_{\mu}}{p\cdot k-i\epsilon}e^{i\vec{k}\cdot \left(\vec{x}-\frac{\vec{p}}{E_{p}}t\right)}+\frac{p_{\mu}}{p\cdot k+i\epsilon}e^{-i\vec{k}\cdot \left(\vec{x}-\frac{\vec{p}}{E_{p}}t\right)}\Bigr].
\end{align}
The requirement 
\begin{align}
~_{out} \langle\!\langle p(t)|\!|A^I_\mu(x) |\!| p(t) \rangle\!\rangle_{out}=A^{adv}_\mu(x;\vp) ~_{out} \langle\!\langle p(t)|\!| p(t) \rangle\!\rangle_{out}
\end{align}
can be satisfied by the following dressing operator
\begin{align}
R_{out}(t)=e\int\! \frac { d ^ { 3 } p } { ( 2 \pi ) ^ { 3 } (2 E_{p}) }\rho(\vec{p})\!\int\! \frac { d ^ { 3 } k } { ( 2 \pi ) ^ { 3 } (2 \omega) }\Big{(}\frac{p^{\mu}}{p\cdot k+i\epsilon}a_{\mu}(\vec{k})e^{i\frac{p\cdot k}{E_{p}}t}-\frac{p^{\mu}}{p\cdot k-i\epsilon}a^{\dagger}_{\mu}(\vec{k})e^{-i\frac{p\cdot k}{E_{p}}t}\Big{)}.
\label{eq:fin_dress}
\end{align}
Thus, the sign of $i\epsilon$ terms is opposite to that in the initial dressing operator $R_{in}$ given by \eqref{eq:ini_dress}.\footnote{This difference of the $i\epsilon$ prescription for initial and final states may be related to the prescription used to define in-out and in-in propagators in nonstationary spacetime \cite{Fukuma:2013mx}.}
This $R_{out}$ is also anti-Hermitian ($R_{out}^\dagger=-R_{out}$), and the dressing factor $e^{-R_{out}}$ is thus unitary. 

The unitarity of the dressing factors, $e^{R_{in}}$ and $e^{-R_{out}}$, guarantees that the asymptotic Hilbert space is positive definite. The asymptotic dressed states are given by multiplying the unitary dressing factors by states satisfying the free Gupta-Bleuler condition. The dressed states thus have a positive norm, because states satisfying the free Gupta-Bleuler condition are positive definite and any unitary transformation preserves the positive definiteness. 

Here, we also give a formal proof of the unitarity of $S$-matrix.
Including the dressing factors, the $S$-matrix acting on the Fock space takes the form (up to phase operators) 
\begin{align}
S=\lim_{t_f\to \infty, t_i\to -\infty} S(t_f,t_i)
\quad \text{with} \quad 
S(t_f,t_i)=e^{-R_{out}(t_f)}S_0(t_f,t_i)e^{R_{in}(t_i)},
\end{align}
where $S_0$ denotes the usual (finite time) $S$-matrix:
\begin{align}
S_0(t_f,t_i)=\mathrm{T}\exp\left(-i\int^{t_f}_{t_i} \!\!\!dt'\, V^I(t')\right) =U^\dagger_0(t_f,t_s)U(t_f,t_i)U_0(t_i,t_s).
\label{eq:Smat0_finite}
\end{align}
The unitarity of $S_0(t_f,t_i)$ simply follows from the expression of eq.\eqref{eq:Smat0_finite}. Since $R_{in}$ and $R_{out}$ are anti-Hermitian, we can show the unitarity of $S(t_f,t_i)$ as
\begin{align}
S^\dagger(t_f,t_i)S(t_f,t_i)&=e^{R_{in}^\dagger(t_i)}S_0^\dagger(t_f,t_i)e^{-R_{out}^\dagger(t_f)}e^{-R_{out}(t_f)}S_0(t_f,t_i)e^{R_{in}(t_i)}\nn
&=e^{-R_{in}(t_i)}S_0^\dagger(t_f,t_i)e^{R_{out}(t_f)}e^{-R_{out}(t_f)}S_0(t_f,t_i)e^{R_{in}(t_i)}
\nn
&=e^{-R_{in}(t_i)}S_0^\dagger(t_f,t_i)S_0(t_f,t_i)e^{R_{in}(t_i)}
\nn
&=1.
\end{align}
Therefore, the $S$-matrix is unitary.

\section{Asymptotic symmetry in the dressed state formalism}\label{sec:asympt}
Gauge theories in 4-dimensional Minkowski space have an infinite number of symmetries (see \cite{Strominger:2017zoo} for a recent review). The symmetries are given by ``large'' gauge transformations such that the gauge parameters can be nonvanishing functions in the asymptotic regions but they preserve the asymptotic behaviors of fields. 
We now discuss the relation between the asymptotic symmetry in QED and dressed states (see also \cite{Mirbabayi:2016axw, Gabai:2016kuf, Kapec:2017tkm, Choi:2017bna, Choi:2017ylo, Carney:2018ygh, Neuenfeld:2018fdw} for related discussions).
We will show that the Faddeev-Kulish dressed states carry the charges associated with the asymptotic symmetry, and investigate the conservation law of the asymptotic charges for the $S$-matrix in the dressed state formalism.

The Lagrangian $\eqref{full-L}$ is invariant under a class of gauge transformations which keep $\partial_\mu A^\mu$ intact. Neother's charge for the transformations in the Schr\"odinger picture is given by 
\begin{align}
Q_{as}^s[\epsilon]=\intx \left[
-\Pi^{0s} \partial_0 \epsilon -\Pi^{is}\partial_i \epsilon+j^{0s}\epsilon
\right],
\label{neother_as}
\end{align}
where the gauge parameter $\epsilon(x)$ satisfies $\Box \epsilon=0$. 
This charge $Q_{as}^s[\epsilon]$ is BRST exact up to the boundary term:  
\begin{align}
Q_{as}^s[\epsilon]=-\intx\, \partial_i (\Pi^{is}\epsilon)
+ \left\{Q^s_{BRST}, \intx(-\bar{c}^s \partial_0 \epsilon+i \pc^s \epsilon)
\right\}.
\label{eq:as_charge-BRST}
\end{align}
Therefore, if the gauge parameter $\epsilon(x)$ vanishes in the asymptotic regions, this charge does not play any role on the physical Hilbert space. However, this is not the case if $\epsilon$ takes nonvanishing values in the asymptotic regions. Such nontrivial charges are called asymptotic charges.  As discussed in \cite{Hirai:2018ijc}, the asymptotic charges are physical charges, \textit{i.e.}, the asymptotic symmetry generated by the charges is not a redundancy of the Hilbert space but the physical symmetry. For example, if $\epsilon$ is a constant, the corresponding charge represents the total electric charge. There is no reason to restrict the Hilbert space to the subspace with zero total electric charge. Similarly, we should not restrict the Hilbert space to the subspace annihilated by $Q_{as}^s[\epsilon]$. Therefore, asymptotic charges $Q_{as}^s[\epsilon]$ can act nontrivially on the physical Hilbert space.

The finiteness condition of $Q_{as}^s[\epsilon]$ requires that  $\epsilon$ should approach functions of angular coordinates around the null infinities. It means that we have an infinite number of asymptotic charges corresponding to the number of functions on two-sphere \cite{He:2014cra}. 
All of the asymptotic charges commute with the BRST charge:
\begin{align}
&[Q_{as}^s, Q^s_{BRST}]=0, 
\end{align}
and they commute with the Hamiltonian $H^s$ up to the BRST exact term:\footnote{This is the reason why we adopted the Hamiltonian \eqref{eq:Hem}. As mentioned in footnote~\ref{foot:ham}, the canonical Hamiltonian $H^s_{can}$ has extra boundary terms: $H^s_{can}=H^s-\int\!\! d^3x\, \partial_i(\Pi_0^s A^{is}+\Pi^{is}A^{0s})$. The boundary terms affect the commutator \eqref{eq:QasH} as $[Q_{as}^s, \int\!\! d^3x\, \partial_i(\Pi_0^s A^{is}+\Pi^{is}A^{0s})]=-i\int\!\! d^3x\, \partial_i(\Pi^{0s}\partial_i \epsilon+\Pi^{is}\partial_0 \epsilon)=\{Q_{BRST}^s,-i\int\!\! d^3x\,\partial_i(\bar{c}^s\partial_i \epsilon)\}-i\int\!\! d^3x\, \partial_i (\Pi^{is}\partial_0 \epsilon)$. Since $\partial_0 \epsilon= \mathcal{O}(r^{-1})$ at $r\to \infty$, we can neglect the effect of boundary terms if the radial component of the electric field operator, $\hat{x}^i\Pi^i$, decays as $\mathcal{O}(r^{-2})$. This condition is probably satisfied for physical scattering states in a reasonable setup.} 
\begin{align}
[Q_{as}^s, H^s]&= -i\intx \left[\partial_i \epsilon \,\partial_i\Pi^{0 s}+\partial_0 \epsilon\, (\partial_i\Pi^{i s}+j^{0s})
\right]
\nn
&=\left\{
Q_{BRST}^s, -i\intx \left(
\partial_i \epsilon \partial_i \bar{c}^s+i\partial_0 \epsilon\, \pc^s
\right)
\right\}.
\label{eq:QasH}
\end{align}
Therefore, the spectrum of the physical Hilbert space is infinitely degenerated. 

This fact naturally leads us to classify the asymptotic states by $Q_{as}^I$ in the interaction picture. 
We now see how $Q_{as}^I$ acts on the initial dressing operator $R_{in}(t)$ given in eq.~\eqref{eq:ini_dress}. As in \eqref{eq:as_charge-BRST}, the asymptotic charge $Q_{as}^I$ in the interaction picture takes the following form up to the BRST exact part: 
\begin{align}
Q_{as}^I[\epsilon]=-\intx\, \partial_i (\Pi^{iI}\epsilon)=-\intx\,[\Pi_i^{I}\partial^i \epsilon+ (\partial_i \Pi^{iI})\epsilon].
\label{eq:as_charge_int_pic}
\end{align}
The commutator of $\Pi_i^{I}$ and $R_{in}(t)$ is given by 
\begin{align}
&[\Pi_i^{I}(t,\vx),R_{in}(t)]\nn
&=ie\int\! \frac { d ^ { 3 } p } { ( 2 \pi ) ^ { 3 } (2 E_{p}) }\rho(\vec{p})\int\! \frac { d ^ { 3 } k } { ( 2 \pi ) ^ { 3 } (2 \omega) }\frac{E_p k_i-\omega p_i}{p\cdot k}\Big{(}e^{-i\vk\cdot(\vx-\frac{\vp}{E_{p}}t)}-e^{i\vk\cdot(\vx-\frac{\vp}{E_{p}}t)}\Big{)},
\end{align}
where we have set $\epsilon=0$ because the integrand is not singular at $\vk=0$. 
On the other hand, the classical electric field for the classical configuration $A_\mu^{ret}(x;\vp)$ given by \eqref{eq:ini_Coulomb} is computed as 
\begin{align}
\partial_0 A_i^{ret}(x;\vp)-\partial_i A_0^{ret}(x;\vp) 
&=ie\int\! \frac { d ^ { 3 } k } { ( 2 \pi ) ^ { 3 } (2 \omega) }\frac{E_p k_i-\omega p_i}{p\cdot k}\Big{(}e^{-i\vk\cdot(\vx-\frac{\vp}{E_{p}}t)}-e^{i\vk\cdot(\vx-\frac{\vp}{E_{p}}t)}\Big{)}.
\end{align}
where we have also set $\epsilon=0$. Hence, one can say that the commutator of $\Pi^{iI}$ and $R_{in}(t)$ is given by the ``classical operator'' which represents the classical  Li\'enard-Wiechert electric field as  
\begin{align}
[\Pi_i^{I}(t,\vx),R_{in}(t)]=\int\! \frac { d ^ { 3 } p } { ( 2 \pi ) ^ { 3 } (2 E_{p}) }\rho(\vec{p})[\partial_0 A_i^{ret}(x;\vp)-\partial_i A_0^{ret}(x;\vp)] \equiv F^{cl}_{0i}(x).
\end{align}
Similarly, the commutator of $\partial_i\Pi^{iI}$ and $R_{in}(t)$ is given by the classical current as 
\begin{align}
[\partial_i\Pi^{iI}(t,\vx),R_{in}(t)]=-e \intp \rho(\vp) \, \delta^3 (\vx- \vp t/E_p)=-j^0_{cl}(x).
\end{align}

Therefore, the asymptotic charge $Q_{as}^I[\epsilon]$ in \eqref{eq:as_charge_int_pic} acts on $e^{R_{in}}$ as 
\begin{align}
[Q_{as}^I[\epsilon], e^{R_{in}}]=e^{R_{in}}\intx\,[F_{cl}^{0i}\partial_i \epsilon+ j^0_{cl}\epsilon].
\label{eq:commu_asI_eRi}
\end{align}
The integral 
\begin{align}
Q^{cl}_{as}[\epsilon]\equiv \intx\,[F_{cl}^{0i}\partial_i \epsilon+ j^0_{cl}\epsilon]
\end{align}
is in fact the  asymptotic charge operator on the Fock space of charged particles. In the limit $t\to \pm \infty$, the eigenvalues agree with the classical leading hard charges computed in  \cite{Hirai:2018ijc}. The leading hard charges are the contributions to the asymptotic charges from uniformly moving charged particles and their Coulomb-like electric fields. For example, if we take a constant $\epsilon=1$, it gives just a total electric charge as $Q^{cl}_{as}[1]\, b^\dagger(\vp) \ket{0}=  e  b^\dagger(\vp) \ket{0}$. If $\epsilon$ is a nontrivial large gauge parameter, the eigenvalues of  $\lim_{t \to \pm}Q^{cl}_{as}[\epsilon]$ are given by momentum-dependent functionals of  $\epsilon^{0}$ which is an arbitrary function on two-sphere that determines the asymptotic behaviors of $\epsilon$  (see \cite{Hirai:2018ijc} for details). 
Therefore, eq.~\eqref{eq:commu_asI_eRi} represents that charged Fock particles with the dressing operator \eqref{eq:FK_R1} carry the asymptotic charges for the classical free charged particles with their Li\'enard-Wiechert electric fields. This result is natural because the dressing corresponds to creating the Li\'enard-Wiechert potential as we have seen in section \ref{sec:meaning}. 

At the classical level, the conservation of asymptotic charges leads to the electromagnetic memory effect  \cite{Hirai:2018ijc}. Let us see the implication at the quantum level. 
To make our discussion simple, we suppose that the radiation sector is given by eigenstates of asymptotic charges at $t=\pm \infty$; that is, we consider states $\ket{\Lambda_{in}}$ and $\bra{\Lambda_{out}}$ such that they contain only transverse photons and satisfy 
\begin{align}
Q^{I,-}_{as}[\epsilon] \ket{\Lambda_{in}}= \Lambda_{in}[\epsilon^{0}] \ket{\Lambda_{in}}, \quad 
\bra{\Lambda_{out}} Q^{I,+}_{as}[\epsilon]= \bra{\Lambda_{out}} \Lambda_{out}[\epsilon^{0}], 
\label{eq:Qas_eigen_rad}
\end{align}
where $Q^{I,\pm}_{as}[\epsilon]= \lim_{t\to \pm \infty}Q^I_{as}[\epsilon]$, and $\Lambda_{in}$ and $\Lambda_{out}$ are arbitrary (c-number) functionals of $\epsilon^{0}$ to which $\epsilon$ asymptotically approaches. We then prepare the following dressed states by exciting charged particles on $\ket{\Lambda_{in}}$ and $\bra{\Lambda_{out}}$ as 
\begin{align}
\ket{in}=e^{R_{in}(t=-\infty)} \hat{\Psi}^\dagger_{in}\ket{\Lambda_{in}},  \quad \bra{out}=\bra{\Lambda_{out}} \hat{\Psi}_{out} e^{-R_{out}(t=+\infty)},
\end{align}
where $\hat{\Psi}^\dagger_{in}$ is an arbitrary product of creation operators $b^\dagger, d^\dagger$ of charged particles and $\hat{\Psi}_{out}$ is any product of annihilation operators $b, d$.
The asymptotic symmetry implies 
\begin{align}
\bra{out} (Q^{I,+}_{as} S_0-S_0Q^{I,-}_{as} )\ket{in}=0, 
\label{as_sym_Smat}
\end{align}
where $S_0$ is given by \eqref{eq:Smat0_finite} with limits $t_f \to \infty, t_i \to -\infty$. 
From \eqref{eq:commu_asI_eRi} and a similar computation for $e^{-R_{out}}$, we have 
\begin{align}
Q^{I,-}_{as}\ket{in}=(Q^-_H+\Lambda_{in})\ket{in}, \quad
\bra{out}Q^{I,+}_{as}=\bra{out}(Q^+_H+\Lambda_{out}).
\end{align}
Here, $Q^-_H$ and $Q^+_H$ represent the hard charge eigenvalues for the states $\hat{\Psi}^\dagger_{in} \ket{0}$ and $\bra{0}\hat{\Psi}_{out}$ respectively as 
\begin{align}
\left(\lim_{t\to -\infty}Q^{cl}_{as}\right) \hat{\Psi}^\dagger_{in} \ket{0}=Q^-_H \hat{\Psi}^\dagger_{in} \ket{0}, \quad 
\bra{0}\hat{\Psi}_{out} \left(\lim_{t\to\infty}Q^{cl}_{as}\right) =\bra{0}\hat{\Psi}_{out} Q^+_H. 
\end{align}
Thus,  \eqref{as_sym_Smat} becomes 
\begin{align}
(Q^+_H+\Lambda_{out}-Q^-_H-\Lambda_{in})\bra{out}S_0\ket{in}=0.
\end{align}
It means  that the $S$-matrix elements can take non-zero values only when the asymptotic charges conserved, 
\begin{align}
Q^+_H+\Lambda_{out}=Q^-_H+\Lambda_{in},
\label{eq:memory}
\end{align}
between the out states and the in states \cite{Kapec:2017tkm}. 
It also means a quantum analog of the classical memory effect. In a scattering event,  if the hard charges are not conserved $Q^+_H\neq Q^-_H$, there should be a change in the radiation sector $\ket{\Lambda_{in}} \to \ket{\Lambda_{out}}$ so that \eqref{eq:memory} holds for any $\epsilon^0$. Conversely, a change in the radiation sector, $\Lambda_{out}-\Lambda_{in}$, is memorized in the change of the hard charges $Q^+_H-Q^-_H$. 

We here comment on the possibility of other dressing operators. 
The standard Fock vacuum is not the eigenstate of $Q_{as}^I$.\footnote{The asymptotic symmetries for general gauge parameters $\epsilon$ are spontaneously broken in the standard Fock vacuum \cite{He:2014cra}.} Roughly speaking, eigenstates in \eqref{eq:Qas_eigen_rad} consist of clouds of soft photons without charged particles. 
However, the Faddeev-Kulish dressing operator $R(t)$ in \eqref{eq:FK_R1} makes a photon cloud only when there are charged particles. 
Thus, we need other dressing operators than Faddeev-Kulish's in order to prepare eigenstates \eqref{eq:Qas_eigen_rad}.
As we have already argued in subsection~\ref{sec:asymptostates}, the dressing operators are not uniquely fixed from the gauge invariant condition. 
We think that this variety is related to the asymptotic symmetry, and leave it for a future work to classify gauge invariant dressed states in terms of the asymptotic charges.

\section{Conclusion and further discussion}

In this paper, we have shown that the Faddeev-Kulish dressed states can be obtained just from the gauge-invariant condition without solving the asymptotic dynamics. While in the original paper \cite{Kulish:1970ut} it was discussed that the dressing operator $R(t)$ in eq.~\eqref{eq:FK_R1} should be modified, we have found that such a modification is not needed. We have also justified the unmodified dressing factor $e^{R(t)}$ with the $i\epsilon$ prescription by giving the interpretation as the Coulomb fields around charges. 
In addition, we have shown the possibility of other types of gauge-invariant dressed states. 
We have also shown that the Faddeev-Kulish dressed states carry the charges associated with the asymptotic symmetry, and have investigated the conservation law of the asymptotic charges for the $S$-matrix in the dressed state formalism.

We close this section with further discussion and comments on future directions. 

\subsection{Softness of dresses}\label{sec:IRfinite}

The infrared finiteness of the dressed state formalism is based on Chung's analysis \cite{Chung:1965zza}. If we extract soft momentum region $k\sim0$ for the dressing operator \eqref{eq:FK_R1}, the operator takes the form 
\begin{align}
R_\text{soft} \sim \sum e \intp \rho(\vp)\int_\text{soft} \! \frac{d^3 k}{(2\pi)^3(2\omega)} \frac{p^\mu}{p \cdot k}\left[
a_\mu(\vk)
-a_\mu^{\dagger}(\vk)
\right],
\end{align}
because $e^{i \frac{p \cdot k}{E_p}t}\sim 1$ at $k\sim 0$. Roughly, this is the dressing operator used in \cite{Chung:1965zza}. In fact, the behavior of the dressing operator at the non-soft momentum region was not specified in \cite{Chung:1965zza}. 
Since only the soft momentum region is relevant for the proof of the IR finiteness, this simplification may be justified.

However, one may worry that the hard momentum contribution in \eqref{eq:FK_R1} affects the physical observables. 
We can make a rough argument that this is not the case as follows. 
First note that $p \cdot k=-\omega (E_p-\vp\cdot\hat{k})$ can be zero only when $\omega=0$ because $p^\mu$ is an on-shell momentum of a massive particle $(E_p>|\vp|)$.
Then, owing to the oscillating factor $e^{i \frac{p \cdot k}{E_p}t}$, contributions from nonzero momenta $(\omega>0)$ can be ignored in the limit $t \to \pm \infty$. 
This statement can be made more rigorous by using $\epsilon$-inserted dressing operator eq.~\eqref{eq:ini_dress} or eq.~\eqref{eq:fin_dress}. 
We shall use the following identity as a distribution: 
\begin{align}
\lim_{\epsilon \to 0}\lim_{t\to \pm \infty}\frac{e^{i \alpha t}}{\alpha \pm i \epsilon}=\mp i \pi \delta(\alpha).
\end{align}
From the identity, we have 
\begin{align}
&\lim_{\epsilon\to 0}\lim_{t\to -\infty} R_{in}(t)=-\frac{i \pi}{2}\sum e \intp \rho(\vp)\intk \frac{p^\mu}{p \cdot k}\left[
a_\mu(\vk)
+a_\mu^{\dagger}(\vk)
\right]\delta(\omega),
\\
&\lim_{\epsilon\to 0}\lim_{t\to \infty} R_{out}(t)=\frac{i \pi}{2}\sum e \intp \rho(\vp)\intk \frac{p^\mu}{p \cdot k}\left[
a_\mu(\vk)
+a_\mu^{\dagger}(\vk)
\right]\delta(\omega).
\end{align}
Therefore, we can say that only soft photons constitute the dresses in the asymptotic limit $t \to \pm\infty$.

Nevertheless, it may be dangerous to use the above asymptotic limit directly. The $S$-matrix on the Fock space is given by 
\begin{align}
\lim_{t_f\to \infty,\, t_i\to -\infty}
e^{-R_{out}(t_f)} \, \mathrm{T}\exp\left(-i\int^{t_f}_{t_i} \!\!\!dt'\, V^I(t')\right) e^{R_{in}(t_i)}.
\label{eq:fin_S}
\end{align}
Thus, we should first compute the finite time $S$-matrix element and then take the limits $t_f \to \infty$ and $t_i \to -\infty$. In addition, since eq.\eqref{eq:fin_S} probably suffers from an infinitely oscillating phase factor, the phase operator such as \eqref{eq:FK_phi1} might be needed to make the $S$-matrix well-defined. As we said in subsec.~\ref{sec:asymptostates}, the phase operator cannot be determined from the gauge invariance. We would like to report a computation of the $S$-matrix in our dressed state formalism including the determination of the phase operator in future.


\subsection{Other future directions}

We would like to comment on other future directions. 

Mandelstam developed a manifestly gauge-independent formalism of gauge theories \cite{Mandelstam:1962mi, Mandelstam:1968hz}. In the formalism, the dynamical variables of QED are the field strength $F_{\mu\nu}$ and path-dependent charged fields such as 
\begin{align}
\phi(x;\Gamma)\equiv e^{-i e \int^x_\Gamma d\xi^\mu A_\mu(\xi)} \phi(x)
\label{eq:wilsonline}.
\end{align}
Such fields attached with Wilson lines are also considered in the context of the bulk reconstruction in the AdS/CFT correspondence (see e.g. \cite{Heemskerk:2012np, Kabat:2012av, Harlow:2015lma}). 
A similarity between Mandelstam's formalism and the dressed state formalism was discussed in \cite{Jakob:1990zi}. However, the dressing operator constructed in \cite{Jakob:1990zi} 
has additional terms depending on the choice of the path $\Gamma$. Thus, the dressing operator seems not to be related directly to Faddeev-Kulish's one \eqref{eq:FK_R1}. Furthermore, we should also investigate the relation to the asymptotic symmetry. As explained in \cite{Harlow:2018tng} for gravitational theories in AdS, operators like \eqref{eq:wilsonline} are transformed under the asymptotic symmetry, and the behavior of the path $\Gamma$ near the asymptotic boundary is important in determining the transformation law of the symmetry. In \cite{Mandelstam:1962mi, Mandelstam:1968hz}, the behavior of $\Gamma$ near the asymptotic region was not specified. Thus, it is interesting to understand more precisely the relations among Mandelstam's formalism, the dressed state formalism and the asymptotic symmetry \cite{future_work}. 

It is important to extend our analysis to other theories. Although the cancellation of IR divergences in the inclusive method \cite{Bloch:1937pw, Yennie:1961ad} was extended to more general theories \cite{Kinoshita:1962ur, Lee:1964is}, we do not know how to define the IR finite $S$-matrices directly. We think that the dressed state formalism is surely useful for this problem. The dressed state formalism for the perturbative gravity was developed in \cite{Ware:2013zja} (see also 
\cite{Choi:2017bna, Choi:2017ylo}). However, a tensor $c_{\mu\nu}$, which is an analog of a vector $c_\mu$ in \cite{Kulish:1970ut}, was introduced by imposing a free ``gauge invariant'' condition which is a gravitational counterpart of the free Gupta-Bleuler condition \eqref{eq:freeGB}. As in QED, we should impose an appropriate physical condition, and we expect that the tensor $c_{\mu\nu}$ is unnecessary.
Asymptotic symmetries for scalar theories are also studied in \cite{Campiglia:2017dpg, Hamada:2017atr, Campiglia:2017xkp}. It was found recently that the asymptotic symmetry of a massless scalar is related to the gauge symmetry of the two-form field dual to the scalar \cite{Afshar:2018apx, Campiglia:2018see, Francia:2018jtb, Afshar:2018sbq, Henneaux:2018mgn}. It might be possible to construct dressed states in a massless scalar theory from the gauge-invariant condition for the dual two-form field.

\section*{Acknowledgement}
The work of SS is supported in part by the Grant-in-Aid for Japan Society for the Promotion of Science (JSPS) Fellows No.16J01004.

\bibliographystyle{utphys}
\bibliography{ref_dress}
\end{document}